\documentclass[prb]{revtex4}
\usepackage{epsf}
\begin{document}
\title{Supermassive Black Holes from collapsing dark matter Bose-Einstein Condensates}
\author{Patrick Das Gupta}
\email{pdasgupta@physics.du.ac.in}
\affiliation{Department of Physics and Astrophysics, University of Delhi,  Delhi-110 007, India}
\author{   }
\author{Eklavya Thareja}
\email{ethare1@lsu.edu}
\affiliation{Department of Physics and Astronomy, Louisiana State University, Baton Rouge, LA 70803, USA}
\begin{abstract}
Discovery of active galactic nuclei at redshifts $\gtrsim 6$  suggests that    supermassive black holes (SMBHs) formed early on. Growth of the remnants of Population III stars by accretion of matter, both baryonic as well as collisionless dark matter (DM), leading up to formation of SMBHs is a very slow process. Therefore, such  models encounter difficulties in explaining quasars detected at $ z \gtrsim 6$. Furthermore, massive particles  making up collisionless DM not only have so far eluded experimental detection but  they also do not satisfactorily explain gravitational structures on small scales. 

In recent years, there is a surge in research activities concerning cosmological structure formation that involve coherent, ultra-light bosons in a dark fluid-like or fuzzy cold DM state. In this paper, we study  
 collapse of such ultra-light bosonic  halo DM  that are in a  Bose-Einstein condensate (BEC) phase to give rise to SMBHs on dynamical time scales. Time evolution  of such self-gravitating BECs is examined using the Gross-Pitaevskii equation in the framework of time-dependent variational method.   
Comprised of identical dark bosons of mass $m$, BECs can collapse to form  black holes of mass $M_{eff}$ on time scales $\sim 10^8$ yrs provided $ m \ M_{eff} \gtrsim 0.64 \ m^2_{Pl} $.  In particular, 
ultra-light dark bosons of mass $ \sim 10^{-20} \ \mbox{eV}$  can lead to SMBHs with mass $\gtrsim 10^{10} \ M_\odot$ at  $ z \approx 6$. 
Recently observed  radio-galaxies in the ELAIS-N1  deep field with  aligned jets can also possibly be explained if vortices of  a rotating cluster size BEC collapse to form spinning SMBHs with  angular momentum $J \lesssim 3.6 \ n_W \frac {G M^2} {c}$, where $n_W$ and $M$ are the winding number and mass of a vortex, respectively. 
\end{abstract}
\maketitle
\section{INTRODUCTION}
\vskip 2.0 em
Fascinating new data concerning supermassive black holes (SMBHs), dark matter (DM) and dark energy (DE) have been enriching the field of gravitation and cosmology in recent years. SMBHs are  frequently getting   discovered at the centers of  galaxies  that include   
our Milky Way which has a SMBH of mass $\sim 4.6 \times 10^{6} \ M_\odot $ \cite{bec0, bec1, bec2}. 
It is also firmly established that gas accreting SMBHs constitute the  central engines  that power active galactic nuclei (AGNs)\cite{bec3}. Many AGNs,  quasars in particular, have been detected at high redshifts, indicating presence of very massive SMBHs early on. Even at z $\approx 6$, when the universe was only $\sim 10^9$ yrs old, several SMBHs with mass in excess of $\sim 10^9 \ M_\odot $  have been observed  \cite{bec4, bec5, bec6, bec7, bec8, bec9}.
SDSS J010013.02+280225.8, the brightest quasar detected so far at a redshift of $z=6.3$, is estimated to contain a SMBH of mass $\sim 1.2 \times 10^{10} \ M_\odot $. \cite{bec10}

A plethora of evidence for DM have been emerging  from observations related to a wide range of cosmic phenomena like rotation curves in galaxies,  gravitationally bound galaxy  clusters and gravitational lensing caused by them as well as  the detected cosmic microwave background anisotropies. These evidence imply existence of DM conclusively provided gravitation ensues from theory of general relativity (GR).  Study  of DM, therefore, is germane not only to astro-particle physics but also to the fundamental aspects of gravitation. Although cold DM (e.g. weakly interacting massive particles (WIMPs)) along with DE dominated,  homogeneous and isotropic, k=0 cosmological models have been successful on scales much bigger than galactic sizes, they  confront inconsistencies when  observations pertaining to structures at  smaller scales are juxtaposed \cite{bec11, bec12}. Moreover, such models face problems explaining presence of high mass SMBHs inferred from observations of distant quasars. 

In this paper, we consider an alternate scenario in which SMBHs are created on dynamical time scales from  collapse of self-gravitating, ultra-light scalars/pseudoscalars which are in a Bose-Einstein condensate (BEC) phase, constituting a fraction of the  
DM halo associated with galaxies. In section II, we   provide a brief discussion on  DM in the light of current observations. Section III describes the formalism of time dependent variational method to estimate the wavefunction  that describes the evolution of BEC in the mean field approximation from Gross-Pitaevskii equation. Condition for the formation of black holes ensuing from an evolving wavefunction of a self-gravitating BEC corresponding to ultra-light bosons is taken up in section IV. In section V, we  consider the  case of a large number of radio-galaxies discovered recently  in the ELAIS-N1 GMRT deep field with their radio-jets pointing roughly in the same direction. We undertake a simplified analysis to explain the observed alignments from the standpoint of  collapsing, cluster size, rotating  dark boson BEC. Finally, the last section provides  concluding remarks pertaining to our proposed mechanism of SMBH generation.

\section{Gravity of dark Matter from recent observations}

Discovery of DM dates back to Zwicky's 1933 analysis of dispersion velocity estimated from the measured line of sight velocities of galaxies in Coma cluster, in which he argued that unless there is excess unseen matter, it is not possible for the cluster to be a  gravitationally bound, virialized system \cite{bec13}. Just over three decades later, study of rotation curves of baryonic matter in near circular orbits far from  galaxy centers provided clinching evidence for DM if one assumes that Newtonian gravitational dynamics is valid on galactic scales \cite{bec14}. Gravitational lensing, which is a direct outcome of GR, lead 
to strong evidence for the existence of  DM in clusters of galaxies \cite{bec15, bec16}.
 
A seminal investigation, involving 153 galaxies with very diverse masses, shapes, dimensions and baryon-to-DM ratios, has recently discovered a  tight correlation between observed rotation curves and that predicted by the observed baryonic matter distribution assuming no DM whatsoever \cite{bec17}. This  result  implies that either gravitational dynamics mimics MOND \cite{bec18}, deviating considerably from GR predictions, or that DM  exists in a superfluid phase and its phonon-like excitations couple with baryons, maintaining a correlation between their distribution \cite{bec19, bec17}.

From particle physics perspective, analysis of ATLAS experiment at LHC consisting of proton-proton collisions with center of mass energy $\sim $ 13 TeV demonstrates that at 95 \% confidence level no new physics beyond the Standard Model (SM) of particle physics has been observed \cite {bec20}. So,  neutralinos and other SUSY particles predicted by  `beyond SM' theories (e.g. supersymmetric extension of SM) that may   constitute WIMPs \cite{bec15, bec21},  are yet to be confirmed. To add to the problems of DM scenarios that rely on WIMPs,  three recent xenon based experiments have severely constrained their  parameter space, excluding at 90 \% confidence level the case of nucleons interacting with WIMPs of mass $\sim $ 50 GeV with  cross-sections above $10^{- 46}\  {\mbox{cm}}^2 $. \cite{bec22, bec23, bec24}
 
One may also ask: can WIMPs aid small seed black holes (BHs) to become SMBHs? In $\Lambda$CDM models, the mass of  first star-forming clouds, generated at redshifts $\sim 20-30$,  are expected to be $\sim \mbox{few}\  10^2 \ M_\odot$ to $\sim \mbox{few}\  10^3 \ M_\odot$ \cite{bec25}. These clouds collapse gravitationally to form  population III (PopIII) stars \cite{bec26}. UV and high energy radiation from such PopIII  stars can help in re-ionizing the universe. Being massive, the first generation stars  evolve  quickly and eventually explode, leaving behind  BHs as  remnants with mass $\gtrsim 10^2 \ M_\odot$.
Various scenarios have been proposed in which high mass SMBHs  can be generated  from  either  infall of matter into seed BHs  heavier than  $\sim  150\ M_\odot $ or  direct collapse of DM halo   \cite {bec26,bec27,bec28, bec29}. 

However, such models encounter   severe challenges in coming up with  SMBHs of mass $\gtrsim 10^9 \ M_\odot$  when the universe was less than $\sim 10^9$ years old. Rate of infall of baryonic matter onto  BHs is limited by the Eddington luminosity. Eddington limit  arises from the fact that as normal matter fall due to the BH's gravity, they get heated up from the infall as their gravitational potential energy turn into thermal energy,  assisted by viscosity and turbulence. The intense radiation from  hot baryonic matter orbiting close to the event horizon of the BH exerts radiation pressure on the falling  electron-proton plasma, and thereby limit the rate of accretion. 
Several studies have shown that in very early galaxies, accretion of baryonic matter happen at  sub-Eddington rate,  limited not only  by outward radiation pressure ensuing from hot matter near  the central BH  but also by low gas density \cite{bec30, bec31}.   
Furthermore, SMBHs are unlikely to grow to have such heavy masses through accretion of particulate DM like WIMPs, as latter can contribute at the most to 10 \% increase in former's mass
\cite {bec32}.  

On the other hand, if  ultra-light scalar/ pseudoscalar particles like axions make up the DM then one can invoke Bose-Einstein condensation of such particles to  create  self-gravitating BECs on astrophysical scales as well as massive primordial BHs \cite{bec33, bec34, bec35, bec36, bec37, bec38, bec39, bec40, bec41, bec42, bec43}. Axion like particles  are predicted not only by gauge theory of strong interactions from CP symmetry considerations but also by  string theories \cite{bec12}. They are expected to be ultra-light since their tiny masses are protected by  approximate global shift symmetry  $a(x^\mu) \rightarrow a(x^\mu) + \mbox{constant} $, where  $a(x^\mu)$ is the axion field. Endowed with large ($\gtrsim  $ 1 kpc scale) de Broglie wavelengths, axions can play the role of fuzzy DM  \cite {bec12,bec43, bec44, bec45, bec46, bec47, bec48, bec49, bec50, bec51, bec52}.  In addition, recent quantum-hydrodynamical  studies demonstrate  that the ensuing large scale cosmic structures from quantum interference of such coherent waves of ultra-light bosons cannot be statistically differentiated from that emerging out of standard cold DM models  \cite{bec52, bec53, bec54}. 

About six years back, in the core of the galaxy cluster Abell 3827, an ongoing merger of at least four  galaxies with a large  cD galaxy was observed \cite{bec55}. Ring like features that arise often due to strong gravitational lensing effects, caused by DM halo, have also been detected around these objects. Modeling these features using  strong gravitational lensing formalism has revealed that the super-giant cD  galaxy is placed asymmetrically in its  DM halo \cite{bec55}. In a recent work, explanation of the observed offset between ordinary matter and DM for this massive structure in Abell 3827 has been posited in terms of collisional dynamics of solitonic galactic cores formed by axions of mass $\approx 10^{-24}$ eV \cite{bec56}.

A new comprehensive study by Hui et al. has highlighted several shortcomings of the standard $\Lambda$CDM model when its predictions are compared with observations on scales $\lesssim $ 10 kpc \cite{bec12}. Some of the observed discrepancies  are - missing DM around globular clusters, lower fraction of satellite galaxies, no sign of central cusps in the DM-dominated galaxies, etc.  The authors have also pointed out that many of these problems can be circumvented if DM is made up of ultra-light bosons with mass $\sim \mbox{few} \times 10^{-22}$ eV so that when their typical speeds are $\sim$ 10 km/s ($\sim $ internal speeds in the central region of galaxies) the corresponding de Broglie wavelength is $\sim $ few kpc, thus making them behave like a quantum fluid that resist compression due to uncertainty principle.  

\section{Bose-Einstein Condensates and Gross-Pitaevskii Equation}

From section II it follows that we can make a strong case for considering coherent ultra-light dark bosons  as a viable alternative to the standard collisionless DM models for studying gravitational structures of size less  than $\sim $ 1 kpc. 
Just over a decade and a half back, it was realized that if DM in galactic halos is comprised of ultra-light bosons with mass $\lesssim 1$ eV then their occupation numbers in the low energy quantum states are astronomically large so that they can be described by a classical field \cite{bec47, bec12}. Interestingly enough, for precisely such light masses, the critical temperature to undergo Bose-Einstein condensation always exceeds the temperature of the universe since the corresponding thermal de Broglie wavelength is larger than the mean separation  between the identical bosons \cite{bec57, bec58}.

The expanding universe provides a conducive environment for ultra-light dark bosons to be in a BEC state. Therefore, it is plausible that a sizable fraction of halo DM  is in the ground state. Moreover, it has been shown that the thermalization necessary for these axions to keep track of the ground state of  BEC is attained through mutual gravitational interactions \cite{bec51}. Typical speeds of such axions are $\lesssim $ 100 km/s, so that non-relativistic quantum analysis is adequate as far as halos are concerned. Dynamics of the halo size BEC is governed by the time evolution of the condensate wavefunction. 

Now, in the  $ T = 0^\circ  K $ mean field approximation,  evolution of the condensate wavefunction  $\psi (\vec{r},t)$ (normalized to unity)  is  described by the Gross-Pitaevskii equation (GPE)\cite{bec59},
$$ i \hbar {{\partial \psi}\over{\partial t}}  = \bigg [- \frac {\hbar^2}{2m} \nabla ^2    + V_{ext} + N \int {V(\vec{r} - \vec{u}) \vert \psi (\vec{u},t)\vert ^2 d^3u} \bigg ] \psi (\vec{r},t) \eqno(1)$$
where $m$, $V_{ext}(\vec{r})$ and $V (\vec{r})$ are the  mass of the boson, an external  potential energy required to   confine the BEC and the interaction potential energy between two bosons, respectively. The boson-boson interaction energy is given by,   
$$ V(\vec{r} - \vec{u})= \frac {4 \pi \hbar ^2  a_s} {m} \delta ^3 (\vec{r} - \vec{u}) + V_g (\vert \vec{r} - \vec{u} \vert)\  \eqno(2)$$
where $\vec {r}$ and $\vec {u}$ are the position vectors of the two bosons, respectively.

In the R.H.S. of eq.(2), the first term is due to short range contact interaction characterized by the s-wave scattering length $a_s$ while the second corresponds to the gravitational interaction between two bosons. In the present study,  the dynamical evolution of BEC is based on eq.(1), with the interaction potential energy appearing in it being given by eq.(2), so that the collective self-gravity of dark bosons arises out of Newtonian two-body gravitational attraction between each pair of identical bosons.
 
Use of eq.(2) in eq.(1) leads to the following GPE,
$$i \hbar {{\partial \psi}\over{\partial t}} = \bigg [- \frac {\hbar^2}{2m} \nabla ^2    + V_{ext} + N g  \vert \psi (\vec{r},t)\vert ^2 + $$
$$  + N \int {V_g (\vert \vec{r} - \vec{u} \vert) \vert \psi (\vec{u},t)\vert ^2 d^3u} \bigg ] \psi (\vec{r},t)\eqno(3) $$
where,
  $$g \equiv \frac {4 \pi \hbar ^2  a_s} {m} \  .\eqno(4)$$


The GPE of eq.(3) can be derived from extremizing the action  $ S = \int {dt \int {d^3r \ \mathcal{L}}} $,
 where,
 $$ \mathcal{L}= \frac {i \hbar}{2} \bigg \lbrace  \psi {{\partial \psi^*}\over{\partial t}} - \psi^* {{\partial \psi}\over{\partial t}}\bigg \rbrace + \frac {\hbar^2}{2m}   \nabla \psi^*. \nabla \psi + V_{ext}\ \vert \psi \vert ^2 + $$ 
 
 $$ \ \ \ \ \ \ \ + \frac {g N} {2} \vert \psi \vert ^4 + \frac {N} {2} \vert \psi \vert ^2 \int {V_g (\vert \vec{r} - \vec{u} \vert) \vert \psi (\vec{u},t)\vert ^2 d^3u} \eqno(5)$$

For the present purpose, we  consider a scenario in which  popIII like stars form at very high redshifts, $z \gtrsim 20 $, and evolve quickly to give rise to compact remnants with mass $M_0 \gtrsim 150 \ M_\odot $. The potential energy of a dark boson due to  such a central compact remnant plays the role of external potential energy $V_{ext}(r) $ of eq.(1) and is given by, 
 
$$ V_{ext} (r) =  - \frac {G M_0 m}  {r} \eqno(6)$$

On scales larger than $\sim $ 50 kpc, possible gravitational effects  on dark bosons due to stars and other baryonic matter, distributed across the proto-galaxy that has just started forming, are  expected to be much  less than that due to their collective DM self-gravity.

Besides  the possible short range weak forces ensuing from binary collisions characterized by the coupling strength $g$ of eq.(4),  dark bosons interact with each other gravitationally, corresponding to which the  Newtonian potential energy  (denoted by  $V_g$ in eqs.(3) and (5)) is simply,
$$ V_g (\vert \vec{r} - \vec{u} \vert) =  - \frac {G m^2}  {\vert \vec{r} - \vec{u} \vert}\ \ .\eqno(7)$$

 Since it is difficult to obtain exact analytical solutions of eq.(3), we adopt   time-dependent  variational method  to capture the essence of  gravitational dynamics of ultracold dark bosons 
 \cite{bec60, bec61, bec62, bec63, bec64}. This technique  employs   trial wavefunctions containing  time-dependent parameters. After choosing a reasonable trial wavefunction, we    determine its parameters  by demanding that the trial wavefunction extremizes the action for which the Lagrangian density is given by eq.(5).
 
We consider the following normalized trial wavefunction
$$\psi(\vec{r},t)= A(t)\exp{(-r^2/2\sigma^2(t))} \exp {(i B(t) r^2)})\eqno(8)$$
where $A(t)$, $\sigma (t)$ and $B(t)$ represent the amplitude, width and a phase parameter for the macroscopic wavefuntion, respectively.
Because of the condition of normalization,  $A(t)$ and $\sigma (t)$ are related by,

$$\vert A(t) \vert^2= (\sqrt{\pi} \sigma(t))^{-3} \Rightarrow A(t)= (\sqrt{\pi} \sigma(t))^{-3/2} \exp{(i \gamma (t))}\eqno(9)$$
Time evolution of the trial wavefunction describing the dynamics of  ultracold dark bosons is characterized by the changes in  the wavefunction width, $\sigma (t) $, and the function $B (t)$ with time. BEC mass enclosed within a sphere of radius $R$ at time $t$  is given by,
$$M_{bec}(<R, t)= N m \int^R _0 {\vert \psi(r,t)\vert ^2 d^3r}$$
$$\ \ \ \ \ = \frac{ 4 \pi N m} {\pi^{3/2} \sigma^3(t)} \int^R _0{r^2 \exp{(- r^2/\sigma^2(t))} d^3r} \eqno(10)$$
so that the BEC mass confined within the Gaussian width $\sigma (t)$  is,
$$M_{eff} \equiv M_{bec}(< \sigma (t))= \frac{ 4 N m} {\sqrt{\pi}}\sum ^\infty_{k=0} {\frac {(-1)^k}{k! \  (2k+3)}} \cong 0.43 \ N m \eqno(11)$$
which turns out to be independent of time. 

If BEC size, characterized by $\sigma (t) $, decreases beyond a limit then the dark boson mass density  can  become so large that GR effects can no longer be overlooked. Since occupation numbers are high for these ultra-light bosons \cite{bec47},  one can do a full general relativistic analysis, in principle, by  assuming a classical field $a(x^\mu)$ to describe  their state. The relevant dynamical equations are,
$$\frac {1} {\sqrt{-g}} \frac {\partial} {\partial x^\mu} \bigg (\sqrt{-g} g^{\mu \nu} \frac {\partial a(x^\alpha)} {\partial x^\nu} \bigg ) + f(a(x))=0$$
$$G_{\mu \nu}= \frac {8 \pi G} {c^4} T_{\mu \nu}(a(x))$$
where $f(a(x))$, $T_{\mu \nu}(a(x))$ and $G_{\mu \nu}$ are the nonlinear term due to self-interactions, stress-tensor for the bosons and Einstein tensor describing ever increasing warping of space-time as the  boson density  increases, respectively. Such an exercise is beyond the scope of the present paper. Making suitable changes in the equation of state for the boson fluid, some GR effects have been included by Chavanis and Harko in  studies related to  general relativistic BEC stars \cite {bec65, bec66}.

 As we have used a non-relativistic formalism in the present work, we will assume  for simplicity that the dark boson condensate collapses to form a black hole when its width becomes smaller than the Schwarzschild radius so that,
$$\sigma (t) < \frac {2 G M_{eff}} {c^2}\ ,\eqno(12) \ $$ 
where $M_{eff}$ denotes the total mass of dark bosons within the width $\sigma (t)$.


In the variational method formalism, the time evolution of $\psi (r,t)$ is determined from  the stationarity of the action\cite{bec60, bec61}. Substituting the trial wavefunction of eq.(8) in the Lagrangian density of eq.(5), and then integrating the latter over space, we arrive at the following 
 Lagrangian, 

$$L= \int {d^3r \ \mathcal{L}} =  \hbar \dot{\gamma}  + L_{int} + \frac {g N} {4\sqrt{2} \pi^{3/2} \sigma^3} + $$

$$\ \   - \ \frac {2G M_0 m}  {\sqrt{\pi}\  \sigma} + \frac{3}{2} \sigma^2 \bigg [\hbar \dot{B} + \frac{2 \hbar^2}{m} B^2 + \frac{\hbar^2}{2 m \sigma^4} \bigg ] \eqno(13)$$
where the self-gravity term $L_{int}$ is given by,
$$L_{int} \equiv \frac {N} {2} \int {d^3r \vert \psi (\vec{r},t) \vert ^2 \int {V_g (\vert \vec{r} - \vec{u} \vert) \vert \psi (\vec{u},t)\vert ^2 d^3u}} \eqno(14)$$
so that for the standard Newtonian gravity (eq.(7)), the trial wavefunction of eq.(8) leads to,
$$L_{int}  = -\ \frac {N G m^2}  {\sqrt{2\pi}\sigma} \ \ .\eqno(15)$$

Euler-Lagrange equations ensuing from extremizing the action,  $\frac{d}{dt} (\partial L/\partial \dot{q_j}) - (\partial L/\partial q_j)=0$, for j=1 and 2, with $q_1 \equiv B$, $q_2\equiv \sigma $ and $L$ given by eq.(13),   have the following forms,
$$\hbar \dot{B} + \frac{2 \hbar^2}{m} B^2 - \frac{\hbar^2}{2 m \sigma^4} - \frac {g N} {4\sqrt{2} \pi^{3/2} \sigma^5}  = - \frac { G [Nm + 2\sqrt{2} M_0]m} {3\sqrt{2\pi} \sigma^3}\eqno(16)$$
$$\dot {\sigma} = \frac {2 \hbar B} {m} \sigma \eqno(17)$$
The variable $\gamma (t) $ is non-dynamical because its corresponding contribution  to eq.(13) is just a  total derivative term. 
Eqs.(16) and (17) can be combined to give,
$$m \ddot{\sigma} = - \frac{dV_{eff}} {d\sigma}\eqno(18)$$
where,
$$V_{eff}\equiv \frac {\hbar^2}{2 m \sigma^2} - \frac{2}{3 \sqrt{2 \pi}} \frac {G[Nm + 2\sqrt{2} M_0]m}{\sigma} + \frac {g N}{6\sqrt{2} \pi^{3/2} \sigma^3}\eqno(19)$$
and,
$$B=\frac{m}{2 \hbar} \frac {\dot {\sigma}}{\sigma}\eqno(20)$$

\section{Formation of Black Holes}
The first term in $V_{eff}$ is a repulsive term which can be thought of arising out of Heisenberg uncertainty principle that enables  existence of solutions representing stable self-gravitating bosonic astrophysical systems\cite{bec33, bec34, bec38, bec41}. The time evolution of the trial wavefunction can be determined by solving eqs.(18) and (20) by specifying the initial data $\sigma (t_i)$  and $\dot {\sigma} (t_i)$ at time $t_i$ which is equivalent to supplying $\psi (t_i)$ by virtue of eqs.(8) and (20).  

Eq.(18)  leads readily to the first integral, 
$$\frac {1}{2} m {\dot {\sigma}}^2 + V_{eff} = \mbox {constant} \equiv K_0 \Rightarrow \dot {\sigma} = \pm \sqrt{\frac {2} {m} \bigg ( K_0 - V_{eff} \bigg )} \eqno(21)$$
From eq.(21) one can determine the turning points by setting $\dot {\sigma}=0$.

\section*{Case I: $K_0=0$ and $g=0$}
Let us consider a scenario in which  a very large number $N$ of dark bosons with $\sim $ zero momentum and  initially spread over a  very large scale $\sim $ 20 - 30 kpc evolve quantum mechanically. Then, if we take the initial conditions to be $ {\dot {\sigma}}^2 \approx 0$ and $B \lesssim 0$ 
with an initial $\sigma_i \sim $ 25 kpc (where $ V_{eff} \approx 0$), the constant $K_0$ of eq.(21) can be taken to be zero so that,
$$\dot {\sigma} = \ -\  \sqrt{ \frac {4 G[Nm + 2\sqrt{2} M_0]}{3 \sqrt{2 \pi } \sigma} - \frac {\hbar^2} {m^2 \sigma^2} - \frac {g N} {3 \sqrt{2} \pi^{3/2} m \  \sigma^3} }\eqno(22)$$
provided,
$$\frac {4 G[Nm + 2\sqrt{2} M_0]\sigma }{3 \sqrt{2 \pi } } - \frac {\hbar^2} {m^2}\bigg [1 + \frac {2 \sqrt{2} N a_s} {3 \sqrt{\pi}\  \sigma} \bigg ] > 0\ .\eqno(23)$$
where one has used eq.(4) for $g$.   An attractive contact interaction ($a_s< 0$) helps gravity to oppose the repulsive force arising due to quantum theory. Eqs.(22) and (23) lead to,
$$t - t_i = \int^{\sigma_i} _{\sigma (t)} { \sigma \   \bigg (\frac {4 G[Nm + 2\sqrt{2} M_0]\sigma }{3 \sqrt{2 \pi } } - \frac {\hbar^2} {m^2} \bigg \lbrace 1 +  \frac {2 \sqrt{2} N a_s} {3 \sqrt{\pi}\ \sigma} \bigg \rbrace \bigg )^{-1/2} d\sigma } \eqno(24)$$

If there is strictly no short range interaction between the bosons  so that  $a_s= 0$, and if the condition eq.(23) is valid all through, eq.(24) can  be readily integrated to arrive at,
$$t - t_i =  \int^{\sigma_i} _{\sigma (t)} { \sigma   \bigg (\frac {4 G[Nm + 2\sqrt{2} M_0]\sigma }{3 \sqrt{2 \pi } } - \frac {\hbar^2} {m^2}  \bigg )^{-1/2} d\sigma }$$

$$ = \bigg (\frac {2} {3} \bigg )  \bigg (\frac {4 G[Nm + 2\sqrt{2} M_0] }{3 \sqrt{2 \pi } \ \sigma^3_i} \bigg )^{-1/2}  
\bigg [ \bigg ( 1 - \frac {3 \sqrt {2 \pi} \hbar^2} {4 G[Nm + 2\sqrt{2} M_0] m^2  \sigma_i} \bigg )^{3/2} \ - \  
 \bigg ( \frac {\sigma (t)} {\sigma_i} - \frac {3 \sqrt {2 \pi} \hbar^2} {4 G[Nm + 2\sqrt{2} M_0] m^2 \sigma_i} \bigg )^{3/2} \ + $$
 
$$ + \ \frac {9 \sqrt {2 \pi} \hbar^2} {4 G[Nm + 2\sqrt{2} M_0] m^2 \sigma_i} \bigg (\sqrt{ 1 - \frac {3 \sqrt {2 \pi} \hbar^2} {4 G[Nm + 2\sqrt{2} M_0] m^2 \sigma_i}}  \ - \ \sqrt{ \frac {\sigma (t)} {\sigma_i} - \frac {3 \sqrt {2 \pi} \hbar^2} {4 G[Nm + 2\sqrt{2} M_0] m^2 \sigma_i}} \bigg ) \bigg ]\eqno(25)$$
As long as the inequality in eq.(12) is violated,  the above equation entails $\sigma (t)$ to decrease steadily with time till it reaches the turning point,
$$\sigma_{min}= \frac {3 \sqrt {2 \pi} \hbar^2} {4 G[Nm + 2\sqrt{2} M_0] m^2 }\eqno(26)$$
where $V_{eff}(\sigma_{min}) = 0$. After the bounce at the turning point,  $\sigma (t) $ starts increasing again. By imposing the criteria,
$$\sigma_{min} >  \frac {2 G M_{eff}} {c^2}\ \eqno(27)$$
that there is {\bf {no}} black hole formation, we can constrain the dark boson mass $m$ as shown below. Of course, if $\sigma_{min} <  \frac {2 G M_{eff}} {c^2}$, there is no bounce and as per our condition given by eq.(12), the condensate collapses to form a black hole.

 Making use of eqs.(11) and (23) in the condition eq.(27), we find the criteria for {\bf {no}} black hole formation to be,
$$ m < \frac {0.64\ m^2_{Pl}} {M_{eff}} \bigg (1+ \frac {1.22 M_0} {M_{eff}} \bigg )^{-1/2} \ \eqno(28a)$$  
where $ m_{Pl} \equiv \sqrt{\hbar c/ G} $ is the Planck mass.

In other words, in order that the initially contracting dark boson condensate under the influence of self gravity does not undergo a bounce due to the uncertainty principle, and instead collapses to form a black hole because of eq.(12), the dark boson mass has to be larger than,
$$ \frac {0.64\ m^2_{Pl}} {M_{eff}} \bigg (1+ \frac {1.22 M_0} {M_{eff}} \bigg )^{-1/2} \ \eqno(28b)$$   
If one chooses $M_{eff} = 10^{10} \ M_\odot$ and $M_0= 150 \ \ M_\odot$, one finds from eq.(28b), that black holes of mass $\sim M_{eff}$ are formed from dark boson BEC provided,
$$ m \gtrsim 10^{-53} \ \mbox{gm} = 0.56 \times 10^{-20} \ \mbox{eV} \ \eqno(29)$$
From eq.(25), one can estimate the time taken for the width of the condensate to decrease to the value of  Schwarzschild radius. Then,  one finds that  collapse to form a black hole takes place on dynamical  time scale,
$$\tau_{dyn} = \bigg (\frac {2} {3} \bigg )  \bigg (\frac {4 G[Nm + 2\sqrt{2} M_0] }{3 \sqrt{2 \pi } \ \sigma^3_i} \bigg )^{-1/2} \approx  10^8 \  \mbox{yrs} \eqno(30)$$
On the other hand, if one considers $M_{eff} = 10^{12} \ M_\odot$ and $M_0= 150 \ \ M_\odot$, one obtains the condition $m \gtrsim 8.7 \times 10^{-23} $ eV in order that the condensate collapses into a SMBH having mass $10^{12} \ M_\odot$ on a  time scale of  $\tau_{dyn} \approx 10^8 $ yrs. Therefore, in this scenario, formation of SMBHs can happen even when the universe is barely $\sim 10^9 $ yrs old. 
\section*{Case II: $K_0>0$ and $g=0$}
If we consider the initial value of $B$ to be negative and large in magnitude, $ {\dot {\sigma}}^2 $  would be initially large  implying $K_0> 0$. Then, according to eq.(18),  the width of the macroscopic   wavefunction will decrease with time till it reaches the turning point where $\sigma $ is   minimum. From eqs. (19) and (21), this occurs when,
$$\frac {2 K_0 \sigma^3} {m} + \frac {4 G[Nm + 2\sqrt{2} M_0] \sigma^2}{3 \sqrt{2 \pi }} - \frac {\hbar^2 \sigma} {m^2} =0 \ ,\eqno(31)$$
so that the turning point occurs at,
$$\sigma_{min} = \frac { G m [Nm + 2\sqrt{2} M_0]  } { 3 \sqrt {2 \pi} K_0} \bigg [ \sqrt {1 + \frac { 9 \pi \hbar^2 K_0} { G^2 (Nm + 2\sqrt{2} M_0)^2   m^3} } - 1 \bigg ] \eqno(32)$$ 
Since, in the present study,  we are considering ultra-cold bosons and $K_0$ is  the classical analogue of energy for a single boson (eq.(21)), we may express it as,
$$K_0 \equiv \epsilon \ m c^2 \  ,\eqno(33)$$
with $0 < \epsilon  \ll 1$.
Use of eq.(33) makes eq.(32) take the following form,
$$\sigma_{min} = \frac { G  [Nm + 2\sqrt{2} M_0]  } { 3 \sqrt {2 \pi}  c^2 \epsilon} \bigg [ \sqrt {1 + \frac { 9 \pi m^4_{Pl} \epsilon} { (Nm + 2\sqrt{2} M_0)^2   m^2} } - 1 \bigg ] \eqno(34)$$ 
 
The condition of eq.(27) for black holes  {\bf {not}} to form entails,
$$\frac {m^4_{Pl}} {m^2} >  2.47 M^2_{eff} \bigg [1 + \frac {1.22 M_0} {M_{eff}} + 3.23 \epsilon \bigg ] \eqno(35)$$
Therefore, for the formation of a black hole, eq.(35) implies that the mass of the dark boson must satisfy,
$$ m > \frac {0.64 \ m^2_{Pl}} {M_{eff}} \bigg (1  + \frac {1.22 M_0} {M_{eff}} + 3.23 \epsilon \bigg )^{-1/2} \eqno(36)$$

\section*{Case III: $K_0<0$ and $g=0$}
When $K_0 <0$, we may modify eq.(33) to,
$$K_0 = - \epsilon \ m c^2 \ , \eqno(37)$$
where  $0 < \epsilon  \ll 1$. Therefore, to obtain the turning point where $\dot {\sigma} $ vanishes, we use eqs.(19), (21) and (37) to arrive at the quadratic equation,
$$\sigma^2_{min} -  \frac {2 G[Nm + 2\sqrt{2} M_0] \sigma_{min}}{3 \sqrt{2 \pi } c^2 \epsilon} + \frac {\hbar^2} {2 m^2 c^2 \epsilon} =0 \ ,\eqno(38)$$
for which the valid root corresponding to the turning point is,
$$\sigma_{min} = \frac { G  [Nm + 2\sqrt{2} M_0]  } { 3 \sqrt {2 \pi}  c^2 \epsilon} \bigg [ 1 - \sqrt {1 -  \frac { 9 \pi m^4_{Pl} \epsilon} { (Nm + 2\sqrt{2} M_0)^2   m^2} }  \bigg ] \eqno(39)$$
Applying the condition given by eq.(12) for black hole formation to eq.(39), one gets the inequality,
$$ m > \frac {0.64 \ m^2_{Pl}} {M_{eff}} \bigg (1  + \frac {1.22 M_0} {M_{eff}} - 3.23 \epsilon \bigg )^{-1/2} \eqno(40)$$

It is easy to see that one may combine the conditions for a black hole formation given by eqs.(28b), (36) and (40) into a single criteria for all values of $K_0$,

$$ m \ \bigg (1  + \frac {1.22 M_0} {M_{eff}} +  3.23 \frac {K_0} {m c^2} \bigg )^{1/2} > \  \frac {0.64 \ m^2_{Pl}} {M_{eff}}  \eqno(41)$$
In the scenario under consideration, $ M_0 \ll M_{eff}  $ and $\vert K_0 \vert \ll mc^2 $. Therefore, the above condition can, for all practical purposes, be expressed  simply as,
$$ m \ M_{eff} \gtrsim 0.64 \ m^2_{Pl} \ ,  \eqno(42)$$
an inequality that essentially reflects an interplay of black hole formation and  uncertainty principle. 

It is evident from eq.(42) that the lower bound on $m$ can be smaller if the mass $\sim M_{eff}$ of the SMBH formed  is larger.
Hence,   ultra-light scalars/pseudoscalars like axions or dynamical four-form (kind of a gravitational axion) \cite{bec67} with mass $m \lesssim 10^{-23} \ \mbox{eV}$,   not only can play the role of DM and DE but also generate SMBHs of mass $\gtrsim 10^{12} \ M_\odot$. Of course, so far the heaviest SMBH  discovered is only of  mass $1.7 \times 10^{10} \ M_\odot$  lying at the centre of an elliptical galaxy, NGC 1600, at a redshift z $\cong $ 0.0157.    \cite{bec68}

\section{Aligned jets of radio-galaxies in  ELAIS-N1 GMRT deep field and quantized vortices of dark boson condensates}

Radio-galaxies emit copiously at radio frequencies, $\sim$ 30 MHz to about $\sim$ 10 GHz, and constitute a subclass of active galactic nuclei (AGNs). Powerful radio-galaxies display long and linear radio loud jets emanating from compact radio-sources  located often in the centers of  elliptical galaxies. These jets tend to be aligned with the optical minor axes of host galaxies. A typical  jet of a luminous radio-galaxy is remarkably aligned over a wide range of length scales  from $\sim $ tens of parsecs to  $\sim $ few kpc to $\sim $ few Mpc, implying a common origin from a compact central engine.
The central engine responsible for making an AGN outshine its host galaxy is comprised  of a  SMBH undergoing accretion of gaseous matter at  relatively higher rate than that in a normal galaxy \cite{bec3, bec69}.

Most central engine models  assume a thick  accretion disc to form around the  SMBH, so that a large fraction of  differentially rotating baryonic matter of the disc encounters viscous dissipation,  and thereby becomes super hot. 
The disc like configuration follows from the initial orbital angular momentum of gas captured by the SMBH. Large luminosity associated with AGNs is due to high rate of  photons emitted by the hot plasma present in the accretion disc as well as due to the radiation from jets made up of collimated, relativistic outflow of  plasma blobs ejected  almost perpendicular to the plane of the disc from regions close to the BH event horizon.  Thickness of the disc and strong helical magnetic fields (generated by frozen magnetic flux in differentially rotating plasma in the disc) help in maintaining a collimated jet. Relativistic effects and viscous dissipation play  dominant roles in making the central engines  efficient in converting gravitational energy of falling baryons into radiation \cite{bec3, bec69, bec70}.   

In a recent study, statistical analysis of position angles of  observed radio jets of 65 radio-galaxies seen within $\sim $ one square degree area of  ELAIS-N1 GMRT deep field  was carried out\cite{bec71}. If one assumes these radio-galaxies to be independent of each other, one would expect no correlation in the orientation of their jets. Findings of Taylor and Jagannathan, however,   strongly suggest that a significant number of jets are aligned with each other on scales larger than  $\sim 0.5^\circ $ (i.e. $\sim  $ 20 Mpc, assuming their cosmological redshifts  to be $\sim $ 0.9).\cite{bec71} The authors  also showed that if the radio-galaxies are not causally connected to one another, the probability of chance alignment in this case is less than 0.1 \%.

Theoretical models that  successfully explain numerous  observed features associated with AGNs entail the jets to be aligned with the spin direction  of Kerr SMBHs \cite{bec72, bec73, bec74, bec75, bec76, bec77}. Hence,  observed alignment of radio-jets  on scales $\gtrsim $ 20 Mpc would imply   that the spin angular momenta of a significant number of  SMBHs  are oriented more or less in the same direction on scales larger than size of rich galaxy cluster. Previous studies involving N-body numerical simulations based on standard CDM models have shown that galactic halo angular momenta  tend to be aligned along the filaments of the  cosmic large scale structure  \cite{bec78}. But question is whether cold DM models  can explain alignments of BH spins on such large angular scales. 

Since  one can also have self-gravitating cluster size BECs in the expanding universe, such observed alignments could be explained naturally in our framework. Tidal torques due to   cosmic large scale structures, acting on cluster size condensates can impart angular momentum to the latter  \cite{bec79, bec80, bec81, bec82, bec83}. Now, laboratory experiments have already demonstrated conclusively that rotating BECs give rise to formation of numerous vortices, analogous to what is observed in superfluid Helium \cite{bec84, bec85}. So, it is plausible for a  rotating dark boson BEC, having a size  larger than a typical rich cluster of galaxies, to break up into  vortices with their angular momenta pointing in the same direction as former's rotation axis. 
 
Indeed, theoretical studies on rotating galactic halos comprised of dark boson condensates reveal that for a wide range of parameters $m$ and $g$, such halos display formation of vortices \cite{bec86, bec87}.  Since each such vortex carries an angular momentum  $ N n_W \hbar $, where $N$ and $n_W$ are the number of dark bosons and winding number, respectively, associated with the vortex, it is likely  that when it  collapses to form a SMBH, the latter would have a spin angular momentum $J \sim  N n_W \hbar $. In what follows, we make back-of-the-envelope  calculations to estimate $a \equiv \frac {c J} {G M^2_{eff}} $  for a SMBH of mass $M_{eff}$ formed from the collapse of a vortex. 

In sections III and IV, we studied the collapse of a spherically symmetric condensate (eq.(8)) to give rise to a Schwarzschild SMBH. Suppose we assume that the results (eqs.(11) and (41)) derived for a  non-rotating BEC is approximately valid for a collapsing vortex with winding number $n_W$ and consisting of $N$ dark bosons of mass m, then it is easy to see that the spin angular momentum parameter,
$$a \sim    \frac {c \hbar N  n_W} {G M^2_{eff}}
= 2.33\ n_W \frac { m^2_{Pl}} {m \ M_{eff}} \eqno(43)$$
Making use of eq.(41) in eq.(43), we obtain an upper bound on $a$,
$$a \lesssim  3.63\ n_W \bigg (1  + \frac {1.22 M_0} {M_{eff}} +  3.23 \frac {K_0} {m c^2} \bigg )^{1/2} \approx 3.63\ n_W  \  . \eqno(44)$$ 
We may compare the above simplified estimate with the measured values of $a$ that rely on   
broadening of  X-ray emission lines corresponding to neutral and partially ionized iron present in the vicinity of accreting SMBHs as well as  fitting of the continuum part of emission from the accretion discs, as these  are very sensitive to the black hole spin rates \cite{bec88, bec89}. Using Fe emission lines  diagnostics,  the spin  parameter  $a \equiv  cJ/GM^2$  has been estimated to be in excess of $0.84 $ for the SMBH in NGC 1365, a  nearby AGN  at a redshift of 0.00545,  \cite{bec90}  and $ a > 0.89 $ for the SMBH in NGC3783, an AGN at z=0.00933, \cite{bec89} both  at 90 \% confidence level. On the other hand, using a different technique involving fitting the observed accretion disc  optical/UV continuum emission and the soft X-ray excess, a Seyfert galaxy PG1244+026, at a redshift of 0.048, is found to have a spinning black hole with $a < 0.86 $. \cite{bec91} 

However, it is important to bear in mind that spin  of a SMBH evolves with time as it accretes gas and stars that carry angular momentum with them, and hence, the measured value of $a$ will in general be different from the  primordial value \cite{bec92, bec93, bec94}. Nevertheless, eq.(44) tells us that even vortices with winding numbers close to unity can lead to rapidly rotating SMBHs. 
   
\section{Conclusions} 
Lack of experimental support for some of the predictions of cold dark matter scenario compounded with the problem of detection of several SMBHs when the universe was barely billion years old motivates one to consider ultra-light bosons as alternate candidates for dark matter. Their negligible rest mass and low speed correspond  to $\sim $ 10 kpc scale de Broglie wavelength and macroscopically large occupation numbers if they make up galactic halos. Gravitational interactions between them aid in re-thermalizing the  ultra-light bosons so that they can form Bose-Einstein condensates.

Using the framework of non-relativistic Gross-Pitaevskii equation in our work, we found that SMBHs with mass $\sim 10^{10} \ M_\odot$  can be created from the collapse of ultra-light dark boson  condensates formed around remnants of population III stars on time scales of $\sim 10^8$ yrs, if the dark boson mass is $ \gtrsim 10^{-20} $ eV. In order that these dark bosons act both as dark matter and dark energy, their rest mass has to be less than $\sim 10^{-23} $ eV, implying that SMBHs with mass $\gtrsim 10^{12} \ M_\odot$ would be generated from  collapse of halo size BEC.

Possible detection, in the future, of high redshift quasars with SMBHs heavier than $\gtrsim 10^{11} \ M_\odot$ will favor our proposed mechanism of generating SMBHs from self-gravitating condensates of ultra-light bosons. A major limitation of our study is that we have not included  general relativistic  BEC dynamics. Instead, we have used a simple criteria of catastrophic gravitational collapse whenever the size of the Bose condensate decreases below the corresponding Schwarzschild radius. 

Recent observation of large scale alignment of jets coming out of radio-galaxies in ELAIS-N1 GMRT deep field suggests that spins of a statistically significant number of SMBHs responsible for jet orientation display preferential alignment. A natural explanation for such a phenomena, in our framework, would be the formation of Kerr SMBHs   due to collapse of  vortices of a rotating dark boson BEC of size larger than a rich cluster of galaxies. Using a simplified analysis, we showed that the observed upper limits  ($\lesssim 0.9 $, in the case of several nearby AGNs) on the SMBH spin parameter $a$  can  be  explained  provided the associated black holes are created from collapse of dark boson BEC vortices with  low winding numbers. If our proposal is in the right track, future observations (with sensitive radio observatories like square kilometer array) are likely to reveal many more cases of such mutually aligned  radio jets on scales larger than galaxy cluster size.

\vskip 2.0 em
 
\begin{acknowledgments}
PDG thanks Ghanashyam Date,  R. Raghavan and D. P. Roy for helpful comments.
\end{acknowledgments}


\end{document}